\documentstyle[aps,preprint,pra]{revtex}  
\begin{document}                
\draft
\title{\bf Quantum state engineering via unitary transformations}
\author{A. Vidiella-Barranco and J.A. Roversi} 
\address{Instituto de F\'\i sica ``Gleb Wataghin'',
Universidade Estadual de Campinas,
13083-970   Campinas  SP  Brazil}  
\date{\today}
\maketitle
\begin{abstract}    
We construct a Hamiltonian for the generation of arbitrary
pure states of the quantized
electromagnetic field. The proposition is based upon the fact that a unitary 
transformation for the generation of number states has been already found. The
general unitary transformation here obtained, would allow the use of nonlinear 
interactions for the production of pure states. We discuss the applicability of 
this method by giving examples of generation of simple superposition states.
We also compare our Hamiltonian with the one resulting from the interaction of 
trapped ions with two laser fields.             
\end{abstract}
\pacs{42.50.Ct, 42.50.Dv} 

The generation of pure nonclassical states of the quantized electromagnetic 
field is a
central topic in quantum optics. Several schemes have been already proposed,
based either on the {\it non-unitary} collapse of the state vector 
(through atom 
measurements) within micromaser environments \cite{schl,barr} or by using a 
cavity QED {\it unitary} time-dependent interaction \cite{eber}. 
Both approaches involve individual atoms interacting with a single-mode 
cavity field, what would demand extraordinary control in a generation 
experiment. It is therefore interesting to seek for alternative methods 
for the generation of nonclassical light. A significant advance was the 
proposal, for the first time, of a unitary generator for the number state 
$|n\rangle$ by Kilin and Horoshko \cite{kili}. They derived an expression for a 
Hamiltonian $\hat{H}_n$ such that
$\hat{H}_n|0\rangle=|n\rangle$ and $\hat{H}_n|n\rangle=|0\rangle$, i.e., they
obtained a Hamiltonian that generates the state $|n\rangle$ from the
vacuum state $|0\rangle$, but only during particular interaction times
$[t_m=(m+1/2)\pi]$. 
This could be experimentally achieved by pumping 
conveniently prepared nonlinear media, as discussed in \cite{kili}.

In this paper we propose a generalization of Kilin-Horoshko's 
procedure for the generation of an arbitrary pure 
state $|\Psi\rangle$, which for instance, can be expressed as a linear 
superposition of number states 
\begin{equation}
|\Psi\rangle=\sum_{n=0}^M C_n\:|n\rangle, 
\label{eq:state}
\end{equation}
where $M$ is the maximum number of photons in the radiation field. 
For certain interaction times, we assume
the existence of a Hamiltonian $\hat{H}_{|\Psi\rangle}$ such that 
$\hat{H}_{|\Psi\rangle}|0\rangle=|\Psi\rangle$ and 
$\hat{H}_{|\Psi\rangle}|\Psi\rangle=|0\rangle$. 
The Hamiltonian $\hat{H}_{|\Psi\rangle}$ may be written in the following general form
\begin{equation}
\hat{H}_{|\Psi\rangle}=\sum_{n=0}^M C_n\left(|0\rangle\langle n|+|n\rangle\langle 
0|\right)-C_0\sum_{k=0}^M |k\rangle\langle k|+\hat{P}\hat{F}\hat{P}
\label{eq:hamil},
\end{equation}
where $\hat{P}=\hat{I}-\sum_{k=0}^M|k\rangle\langle k|$ and $\hat{F}$ is any 
Hermitian operator. The sums of projectors 
$\sum_{k=0}^M|k\rangle\langle k|$ are needed in order to compensate all
the terms present in the superposition state in Eq.(\ref{eq:state}).
For the sake of simplicity we have considered the coefficients 
$C_n$ real. Our task now is to find an operator $\hat{F}$ such that we
are able to delete the higher order powers of $\hat{a}$ and $\hat{a}^\dagger$
in order to obtain the simplest possible expression for $\hat{H}_{|\Psi\rangle}$.
We start by decomposing $\hat{F}$ in a convenient way
\begin{equation}
\hat{F}(\hat{a},\hat{a}^\dagger)=\hat{f}_0(\hat{a}^\dagger\hat{a})+\sum_{m=1}^M
\left[\hat{f}_m(\hat{a}^\dagger\hat{a})\frac{\hat{a}^m}{\sqrt{m!}}+
\frac{\hat{a}^\dagger{}^m}{\sqrt{m!}}\hat{f}^*_m(\hat{a}^\dagger\hat{a})\right].
\label{eq:efe}
\end{equation}
Now if we substitute Eq.(\ref{eq:efe}) and the expression for $\hat{P}$ into
Eq.(\ref{eq:hamil}), after some algebra we find the conditions that the 
functions $f_l(k)\equiv \langle k|\hat{f_l}(\hat{a}^\dagger\hat{a})|k\rangle$ 
must obey in order to cancel most of the terms in $\hat{H}_{|\Psi\rangle}$.
These conditions are
\begin{equation}
f_0(0)=C_0; \ \ \ \ \ \ f_0(k)=-C_0; \ \ \ \ \ \ f_m(0)=C_m ; \ \ \ \ \ \ 
f_m(k)=0, \label{eq:cond}
\end{equation}
for $k=1,2, \ldots M$ and $m=1,2, \ldots M$. Those are necessary but not 
sufficient conditions for the determination of the functions 
$\hat{f}_m(\hat{a}^\dagger\hat{a})$, what introduces a 
certain degree of arbitrarieness in their choice. A simple form for 
$\hat{f}_0$ and $\hat{f}_m$ involving a {\it finite expansion\/} in the 
annihilation and creation operators $\hat{a}$ and $\hat{a}^\dagger$ is
\begin{equation}
\hat{f}_0(\hat{a}^\dagger\hat{a})=C_0\left[2\left(1-\hat{a}^\dagger\hat{a}
\right) \,{\cal {F}}(\hat{a}^\dagger\hat{a})-1\right];\ \ \ \ \ \ \ \
\hat{f}_m(\hat{a}^\dagger\hat{a})=C_m\:{\cal {F}}(\hat{a}^\dagger\hat{a});
\ \ \ \ \ \ \ \ m\neq 0,\label{eq:efes}
\end{equation}
where
\begin{equation}
{\cal {F}}(\hat{a}^\dagger\hat{a})=\sum_{l=0}^M
A_l\:(\hat{a}^\dagger\hat{a})^l.
\end{equation}
The condition $f_m(0)=C_m$ demands that 
${\cal {F}}(\hat{a}^\dagger\hat{a})|0\rangle=|0\rangle$, 
which means that we must have $A_0=1$. 
Moreover, we can use the condition $f_m(k)=0$ [see 
Eq.(\ref{eq:cond})] to determine the remaining coefficients $A_l$. The 
successive application of the function 
${\cal {F}}(\hat{a}^\dagger\hat{a})$ onto the $M$ number states $|1\rangle$,
$|2\rangle\ldots |M\rangle$ gives rise to the following $M$ coupled linear 
equations for the coefficients $A_l$: 
 
\begin{eqnarray}
1&+& A_1+A_2+\ldots +A_M=0 \nonumber\\
1&+& 2A_1+4A_2+\ldots +2^{M}A_M=0\nonumber\\
\vdots \nonumber\\
1&+& MA_1+M^2 A_2+\ldots +M^M A_M=0.\label{eq:siseq}
\end{eqnarray}
The solution of this set of equations always exists, what 
completely determines the functions $\hat{f}_m$. 
Therefore there is a specific set of numerical coefficients $A_l$ for every 
value of $M$, and which are of the form: $A_1=\alpha_1/M!, 
A_2=\alpha_2/M!,\ldots A_M=\alpha_M/M!$, with 
$|\alpha_1|<|\alpha_2|<\ldots<|\alpha_{M-1}|<|\alpha_M|$. 
In particular, $\alpha_M=(-1)^M/M!$. This result is convenient for us 
because the powers of
$\hat{a}^\dagger\hat{a}$ will be multiplied by increasingly smaller
coefficients, i.e., the relative importance of the higher-order terms will
be consistently diminished.

The complete and already simplified Hamiltonian will then read
\begin{equation}
\hat{H}_{|\Psi\rangle}=f_0\left(\hat{a}^\dagger\hat{a}\right)+
\sum_{m=1}^M\frac{C_m}{\sqrt{m!}}
\left[{\cal {F}}(\hat{a}^\dagger\hat{a})\,
\hat{a}^m+\hat{a}^\dagger{}^m\,{\cal {F}}(\hat{a}^\dagger\hat{a})\right].
\label{eq:hamilf}
\end{equation}
We would like to remark that the function $f_0(\hat{a}^\dagger\hat{a})$ is
important for the establishment of the match between our Hamiltonian and
the ``physical'' interaction Hamiltonian in a realistic experiment.
Here, we have made a choice for $f_0$ [see Eq.(\ref{eq:efes})] that allows the
inclusion of the Kerr effect in the generation process, as we are going to show.
 
In what follows we discuss applications of this method. It would be
interesting to generate a state exhibiting nonclassical properties such as 
squeezing, antibunching and sub-Poissonian character, for instance. These
effects occur simultaneously in a single state, namely the binomial state
\cite{sto}, which admits an expansion as the one in 
Eq.(\ref{eq:state}), having a finite number of number state (real) coefficients
different from zero
\begin{equation}
C_n=B_n^M=\left[\frac{M!}{n!\,(M-n)!}\, p^n (1-p)^{M-n}\right]^{1/2}.
\end{equation} 
Its photon number distribution $P_n=(B_n^M)^2$ is a binomial 
distribution. The binomial states are caracterized by two parameters: $p$ being 
the probability of emission of a single photon, and $M$ the maximum number of 
photons in the field. They interpolate between a number state 
$|M\rangle$ (containing $M$ photons), as $p\rightarrow 1$, and a coherent state 
$|\alpha\rangle$ (with real amplitude $\alpha=\sqrt{pM}$)
as $p\rightarrow 0$ and $M\rightarrow\infty$, i.e.,
they belong to a class of ``intermediate states''. Their nonclassical 
properties, of course, are strongly dependent on the values of the 
``interpolation parameters'' $p$ and $M$ \cite{avb1}. Generalizations of binomial
states also include the squeezed coherent states \cite{sasa}.

We start by showing an explicit calculation of the Hamiltonian for the 
particular case of $M=1$, i.e., the state being generated would be the 
superposition of the vacuum state
with the one-photon state $|\phi\rangle=C_0|0\rangle+C_1|1\rangle$. 
In this case we have that 
${\cal {F}}(\hat{a}^\dagger\hat{a})=1+A_1\hat{a}^\dagger\hat{a}$. From the
requirement $(1+A_1\hat{a}^\dagger\hat{a})|1\rangle=0$ [fourth condition in
Eq.(\ref{eq:cond})], we obtain
$A_1=-1$. The Hamiltonian in Eq.(\ref{eq:hamilf}) then becomes  
\begin{equation}
\hat{H}_{|\phi\rangle}=(1-p)^{1/2}\left[1-4\hat{a}^\dagger\hat{a}+2
\left(\hat{a}^\dagger\hat{a}\right)^2\right]+p^{1/2}\left[
(1-\hat{a}^\dagger\hat{a})\hat{a}+\hat{a}^\dagger(1-\hat{a}^\dagger\hat{a})
\right].\label{eq:hamils}
\end{equation}

On the other hand, if we pump with a classical field 
$E_c=Ee^{-i\Omega t}+E^*e^{i\Omega t}$ (polarized along the $y$ direction)
a nonlinear medium characterized by linear and nonlinear susceptibilities 
$\chi^{(1)}$ and $\chi^{(3)}$ respectively, the coupling of the signal
(polarized along the $x$ direction), the pump and the output fields will be 
described by the following Hamiltonian \cite{kili}:
\begin{eqnarray}
\hat{H}_{NL}=\chi_{xx}^{(1)}\hat{a}^\dagger\hat{a}+\chi_{yy}^{(1)}|E|^2+
+\chi^{(3)}_{xxxx}\hat{a}^\dagger{}^2\hat{a}^2
+\chi^{(3)}_{xyxy}\hat{a}^\dagger 
\hat{a}|E|^2+\chi^{(3)}_{yyyy}|E|^2|E|^2\nonumber\\
+\left(\chi_{xy}^{(1)}\hat{a}^\dagger E+ 
\chi^{(3)}_{xxyy}\hat{a}^\dagger{}^2E^2+\chi^{(3)}_{xxxy}\hat{a}^\dagger{}^2
\hat{a}E+\chi^{(3)}_{xyyy}\hat{a}^\dagger |E|^2E+\mbox{H.c.}\right).
\label{eq:hamilnl}
\end{eqnarray}
Both the pump and signal are travelling waves propagating along the $z$ axis.
The expression for our generator [in Eq.(\ref{eq:hamils})] has the same form as
the Hamiltonian in Eq.(\ref{eq:hamilnl}), what allows, at least in 
principle, to know how to choose and prepare a nonlinear medium in such a
way that we end up with the Hamiltonian in Eq.(\ref{eq:hamils}), as we are going
to show. The linear and nonlinear
susceptibilities have to assume specific values in order to make possible the
correspondence between both Hamiltonians. 
For instance, there is no term in Eq.(\ref{eq:hamils}) 
proportional to $\hat{a}^\dagger{}^2$, which means that we must have
$\chi^{(3)}_{xxyy}=0$. By comparing terms proportional to $\hat{a}^\dagger$,
$\hat{a}^\dagger{}^2\hat{a}$ and to $\hat{a}^\dagger\hat{a}$, we obtain
\begin{equation}
\chi_{xy}^{(1)}(E_0)E+\chi^{(3)}_{xyyy}|E|^2E=-\chi^{(3)}_{xxxy}E=p^{1/2},
\label{eq:chixy}
\end{equation}
and
\begin{equation}
\chi_{xx}^{(1)}(E_0)+\chi^{(3)}_{xyxy}|E|^2=-\chi^{(3)}_{xxxx}=-2(1-p)^{1/2}.
\label{eq:chixx}
\end{equation}
These relations among the different susceptibilities are of the same form as the
ones found in reference \cite{kili}. However, in our case we are generating a state
which is the one-photon state coherently superposed to the vacuum, fact
which is embodied in the ``tuning'' parameter $p$. The susceptibilities 
$\chi^{(3)}_{ijkl}$ are of course a feature of the chosen crystal and the 
first-order susceptibilities $\chi_{ij}^{(1)}$ 
can be changed by the application of a static field $E_0$. 
We can control the generation procedure by adjusting the values 
of the pump field $E$ and the static field $E_0$ in order to satisfy the relations 
(\ref{eq:chixy}) and (\ref{eq:chixx}). From Eq.(\ref{eq:chixy}) we see that 
the probability of having one photon in the field, $p$, is proportional to 
the pump field amplitude $E$, as one would expect. We note that the conditions in
Eq.(\ref{eq:chixy}) and (\ref{eq:chixx}) connect the quantum superposition
principle, represented by the coherent superposition of the vacuum with the one-photon
state, with ``macroscopic features'' such as nonlinear susceptibilities in a crystal. 

Our general scheme allows the generation, in principle, of virtually 
any pure state of the quantized field through some kind of nonlinear 
interaction. As a second example, we now construct the generating Hamiltonian for 
the superposition of the vacuum state 
$|0\rangle$ with the two-photon state $|2\rangle$, or $|\psi\rangle=C_0|0\rangle+
C_2|2\rangle$. In this case the solution of the system of
equations in Eq.(\ref{eq:siseq}) gives us $A_1=-3/2$, $A_2=1/2$, and the Hamiltonian
will read   
\begin{eqnarray}
&&\hat{H}_{|\psi\rangle}=C_0\left[1-5(\hat{a}^\dagger\hat{a})+4
(\hat{a}^\dagger\hat{a})^2-1(\hat{a}^\dagger\hat{a})^3\right]
+\frac{C_2}{\sqrt{2}}\left[{\cal F}(\hat{a}^\dagger\hat{a})\hat{a}^2
+\hat{a}^\dagger{}^2{\cal F}(\hat{a}^\dagger\hat{a})\right];\nonumber \\ 
&&{\cal F}(\hat{a}^\dagger\hat{a})=1-\frac{3}{2}
\hat{a}^\dagger\hat{a}+\frac{1}{2}(\hat{a}^\dagger\hat{a})^2.\label{eq:hatwo}
\end{eqnarray} 
We note that because the target state $|\psi\rangle$ above does
contain even photon numbers only, there are fewer terms in the generating 
Hamiltonian, what means that additional conditions must be imposed on the 
relevant susceptibilities in the corresponding nonlinear Hamiltonian. More
specifically, in Eq.(\ref{eq:hatwo}) we have terms of the type 
$\hat{a}^\dagger{}^2\left(\hat{a}^\dagger\hat{a}\right)^2$, which means that  
a fourth-order nonlinear susceptibility ($\chi^{(4)}$) should take part in the 
generation process. Recently there have been developments both theoretical 
\cite{davi} and experimental \cite{balak} in processes involving five-wave 
mixing in fluids. This kind of medium has several advantages, such as a 
flexible geometry, for instance. However, due to its 
intrinsic isotropy, processes involving modes in co-linear propagation are 
forbidden. In general there are more stringent conditions over fluid media than in
crystals. This of course may favour our scheme in the sense that 
we have only a few terms present in our Hamiltonian.

We would expect that the Hamiltonian in Eq.(\ref{eq:hamilf}) for the generation 
of binomial states (in the limits of $p\rightarrow 0$ and $M\rightarrow\infty$)
should be somehow equivalent to Glauber's displacement 
operator $\hat{D}(\alpha)=\exp(\alpha\hat{a}^\dagger-\alpha^*\hat{a})$. 
In fact we have
\begin{eqnarray}
\hat{H}_{|\Psi\rangle\rightarrow|\alpha\rangle}|0\rangle&=&\lim_{p\rightarrow 0, 
M\rightarrow\infty}\left[B_0^M|0\rangle+\sum_{m=1}^\infty 
B_m^M\frac{\hat{a}^\dagger{}^m}{\sqrt{m!}}|0\rangle\right]=\sum_{m=0}^\infty
\frac{e^{-\alpha^2/2}\alpha^m\hat{a}^\dagger{}^m}{m!}|0\rangle\nonumber\\
& &\nonumber\\  
&=&e^{-\alpha^2/2}e^{\alpha\hat{a}^\dagger}|0\rangle=
e^{-\alpha^2/2}e^{\alpha\hat{a}^\dagger}e^{\alpha^*\hat{a}}|0\rangle=
\hat{D}(\alpha)|0\rangle=|\alpha\rangle,
\end{eqnarray}
i.e., the application of either the appropriate Hamiltonian $\hat{H}_{|\Psi\rangle}$ 
or $\hat{D}(\alpha)$ onto the vacuum state leads to the same (coherent) state
$|\alpha\rangle$. 
It is also worth verifying that the Hamiltonian in Eq.(\ref{eq:hamilf}) 
(in the limit of $p\rightarrow 1$) is equivalent 
to Kilin-Horoshko's Hamiltonian \cite{kili} for the generation on number
states. In this case $B_n^M=\delta_{n,M}$, and
\begin{equation}
\hat{H}_{|\Psi\rangle\rightarrow|M\rangle}=\frac{1}{\sqrt{M!}}
\left[{\cal {F}}(\hat{a}^\dagger\hat{a})
\hat{a}^M+\hat{a}^\dagger{}^M{\cal {F}}(\hat{a}^\dagger\hat{a})\right],
\end{equation}
so that
\begin{equation}
\hat{H}_{|M\rangle}|0\rangle=\frac{1}{\sqrt{M!}}\hat{a}^\dagger{}^M
\left[1+\sum_{l=1}^M A_l(\hat{a}^\dagger\hat{a})^l\right]|0\rangle=|M\rangle.
\end{equation}
The difference between Kilin-Horoshko's Hamiltonian and ours with respect to
the number state case rests on the functions 
${\cal {F}}(\hat{a}^\dagger\hat{a})$. They do not need to contain higher 
order powers of $(\hat{a}^\dagger\hat{a})$. In fact a simpler function  
${\cal {F}}=1-\hat{a}^\dagger\hat{a}/M$ is enough for the number state 
generation \cite{kili}. However, we shall stress that in a real experiment,
a nonlinear medium has
in general all its nonlinear susceptibilities excited as it is pumped. This 
means that it is convenient to have the contributions of all powers of
$\hat{a}^\dagger\hat{a}$ (up to the $M$th) in the interaction Hamiltonian, 
as it does happen in our approach. Our generalization, then, is consistent 
with previously known results. 

The effective implementation of this technique would constitute a 
challenging experimental problem, starting from the design of appropriate 
media. It would be interesting a comment on an alternative physical system, 
other than 
the electromagnetic field, that could be suitable for the accomodation of such a 
nonlinear Hamiltonian. Recently there has been a great deal of interest in the 
generation of nonclassical states of motion of a single ion confined into an 
electromagnetic trap. 
There have been carried out successful experiments for the production of number, 
coherent and squeezed vacuum states of the ion's motion \cite{monro}. 
On the other hand, Raman-type excitation of a trapped ion via two laser fields 
produces an intrinsically nonlinear Hamiltonian, which is conveniently 
expressed (one-dimensional case), in the following form 
\cite{vogel}:
\begin{equation}
\hat{H}_{ion}=\frac{1}{2}\hbar\Omega\hat{f}_k(\hat{n};\eta) 
(i\eta\hat{a})^{k}+ H.c.; \ \ \ \ \ \ \mbox{where} \ \ \ \ \ \
\hat{f}_k(\hat{n};\eta)=e^{-\eta^2/2}\sum_{l=0}^\infty\frac{(-1)^l\eta^{2l}}
{l!(l+k)!}\;\hat{a}^\dagger{}^l\hat{a}^l.
\label{eq:havog}
\end{equation}
Here $\eta=2\pi a_0/\lambda$ is the Lamb-Dicke parameter, being 
$a_0(\equiv 1/\sqrt{2m\nu_1})$ the size of the ground state of the 
harmonic potential of natural frequency $\nu_1$, 
$\Omega=\Omega_1\Omega^*_2/2(\omega_{21}-\omega_L)$ 
is the system's effective two-photon Rabi frequency, $\Omega_i$ are the
single-photon Rabi frequencies, $\omega_{21}$ the atomic transition
frequency, and $\omega_L$ is the laser frequency. 
The detuning between the lasers is $\Delta=k\nu_1$ ($k=1,2,\ldots$). 
This decomposition in Eq.(\ref{eq:havog}) is similar to the one
we used in our Hamiltonian [Eq.(\ref{eq:hamilf})], the difference being that
in the ion's case the functions $\hat{f}_k(\hat{n})$ are already fully 
specified. For instance, if we consider $\eta\ll 1$, 
and $k=1$, we may write
\begin{equation}
\hat{f}_1(\hat{n};\eta)\approx e^{-\eta^2/2}\left(1-\frac{\eta^2}{2}
\hat{a}^\dagger\hat{a}\right),
\end{equation}
and the interaction Hamiltonian will read
\begin{equation}
\hat{H}_{ion}=\frac{1}{2}i\hbar\eta e^{-\eta^2/2}\left[
\Omega\left(1-\frac{\eta^2}{2}\hat{a}^\dagger\hat{a}\right)\hat{a}-
\Omega^*\hat{a}^\dagger\left(1-\frac{\eta^2}{2}\hat{a}^\dagger\hat{a}\right)
\right].
\end{equation} 
The Hamiltonian above has the same form as the one in Eq.(\ref{eq:hamils}), for the 
generation of the one-photon state ($p\rightarrow 1$). 
We note that despite the similarity between both Hamiltonians, they are not totally
equivalent, because for that we must have $\eta=\sqrt{2}$, which is in
contradiction with our assumption $\eta\ll 1$, apart form other differences.
Therefore it is not so evident the correspondence between the quantum state
engineering we are here proposing, and highly nonlinear systems such as driven
trapped ions. Nevertheless, the approach adopted here,  based on fixed 
interaction times, should be in our opinion further investigated in order 
to seek for a clearer connection between our tailor-made Hamiltonian and a 
scheme of excitation of ions by laser beams.

We have presented here the explicit construction of a Hamiltonian 
$\hat{H}_{|\Psi\rangle}$
which is a generator of nonclassical states of the quantized 
electromagnetic field. This extends the class of methods being considered 
for this purpose, drawing together, as particular cases, previously known 
generators such as the number state as well as the coherent state ones.
Having a generation scheme based on unitary transformations in hands, 
new possibilities arise for the engineering of nonclassical light under 
feasible controlled circumstances, e.g., through nonlinear interactions. 
Moreover, this scheme has the advantage of involving travelling waves, what 
obviously minimizes the destructive effect of dissipation.
 
\acknowledgments

We would like to thank M. Marchiolli and one of the referees for valuable comments.
This work was partially supported by CNPq (Conselho Nacional para o 
Desenvolvimento Cient\'\i fico e Tecnol\'ogico, Brazil).


\begin{references}

\bibitem{schl} K. Vogel, V.M. Akulin and W.P. Schleich, Phys. Rev. Lett. {\bf 
71,} 1816 (1993).

\bibitem{barr} B.M. Garraway, B. Sherman, H. Moya-Cessa, P.L. Knight and G. 
Kurizki, Phys. Rev. A {\bf 49,} 535 (1994).

\bibitem{eber} C.K. Law and J.H. Eberly, Phys. Rev. Lett. {\bf 76,} 1055 (1996).

\bibitem{kili} S. Kilin and D. Horoshko, Phys. Rev. Lett. {\bf 74,} 5206 (1995).

\bibitem{sto} D. Stoler, B.E.A. Saleh and M.C. Teich, Optica Acta {\bf 32,} 
345 (1985).

\bibitem{avb1} A. Vidiella-Barranco and J.A. Roversi, Phys. Rev. A {\bf 50,} 
5233 (1994).

\bibitem{sasa} H-C. Fu and R. Sasaki, J. Phys. A {\bf 29,} 5637 (1996).

\bibitem{davi} L.C. D\'avila Romero, S.R. Meech, and D.L. Andrews, J. Phys. A {\bf 30},
5609 (1997).

\bibitem{balak} A.V. Balakin, N.I. Koroteev, A. Pakulev, A.P. Shkurinov, D. Boucher,
P. Masselin, and E. Fertein, JETP Lett. {\bf 64}, 718 (1996).

\bibitem{monro} D.M. Meekhof, C. Monroe, B.E. King, W.M. Itano and
D.J. Wineland, Phys. Rev. Lett. {\bf 76,} 1796 (1996).

\bibitem{vogel} S.Wallentowitz and W.Vogel, Phys. Rev. A {\bf 55,} 4438 (1997).
  

\end{references}
\end{document}